\documentclass[11pt]{article}

\usepackage[utf8]{inputenc}
\usepackage[T1]{fontenc}
\usepackage{lmodern}
\usepackage{mathtools}
\usepackage[margin=1in]{geometry}
\usepackage{amsmath, amssymb}
\usepackage[super,sort&compress]{natbib}
\usepackage{graphicx}
\usepackage{booktabs}
\usepackage{subcaption}
\usepackage{tabularx}

\usepackage{siunitx}
\usepackage{caption}

\usepackage{placeins}   
\usepackage{needspace}  
\usepackage{hyperref}

\usepackage{comment}
\usepackage{authblk}

\hypersetup{colorlinks=true, linkcolor=black, citecolor=black, urlcolor=blue}

\title{ Bayesian Sequential Modeling of Time-to-Urination for Dynamic ED Triage}
\author[1,2]{Atsushi Senda\thanks{Corresponding author: sendaccm@tmd.ac.jp https://orcid.org/0000-0002-0128-6800
}}
\author[2] {Yuki Takatsu}
\author[3]{Ryokan Ikebe}
\author[2]{Hiroshi Suginaka}
\author[1]{Koji Morishita}
\author[4]{Akira Endo}

\affil[1]{Department of Acute Critical Care and Disaster Medicine, Graduate School of Medical and Dental Sciences, Institute of Science Tokyo, Tokyo,  Japan}
\affil[2]{Department of Emergency Medicine, Toda Chuo General Hospital, Toda, Saitama, Japan}
\affil[3]{Department of Emergency Medicine, Tokyo Women's Medical University, Tokyo, Japan}
\affil[4]{Department of Acute Critical Care Medicine, Tsuchiura Kyodo General Hospital: Tsuchiura, Japan}

\date{\today}

\begin{document}
\maketitle

\begin{abstract}
Triage tools in routine emergency care are largely static, failing to exploit simple behavioral cues clinicians notice in real time. Here, we developed a Bayesian, sequentially updating framework that integrates incoming cues to produce calibrated, time-consistent risk. Using a prospective single-center cohort of ambulance arrivals in Japan (February–August 2025; n=2,221), we evaluated time to first urination (TTU) as a proof-of-concept bedside cue for predicting hospital admission. Population-level fit to the cumulative admission curve was excellent (integrated squared error 0.002; RMSE 0.003; Kolmogorov–Smirnov 0.008; coverage 0.98). At the patient level, performance improved markedly with age/sex adjustment (AUC(t) 0.70 vs. 0.50 unadjusted), with lower Brier scores and positive calibration slopes. Platt recalibration refined probability scaling without altering discrimination, and decision-curve analysis showed small, favorable net benefit at common thresholds. This framework is readily extensible to multimodal inputs and external validation and is designed to complement, not replace, existing triage systems.
\end{abstract}

\section{Introduction}
Triage in emergency medicine remains limited by static designs that rely on single-time-point assessments and overlook dynamic, non-verbal cues clinicians intuitively recognize at the bedside~\citep{Worster2005, vanderWulp2009, StormVersloot2011}. Although artificial intelligence (AI) such as convolutional neural networks for image classification/detection and segmentation---has achieved notable success in radiology and pathology, fueled by large datasets and well-defined tasks that enable robust models~\citep{Esteva2017, Rajpurkar2017, Campanella2019}, its triage application has been limited. Real-world evaluations continue to expose performance gaps and implementation challenges in diagnostic reasoning and triage~\citep{Levin2018, Raita2019, Peck2013}. Conversely, clinical reasoning and triage are inherently dynamic, evolving as new information emerges~\citep{Churpek2016, Kipnis2016, Escobar2020}. Temporal features such as vital-sign trajectories improve recognition of deterioration compared with snapshot assessments~\citep{Churpek2016Resuscitation, Kipnis2018}. Moreover, exi
sting tools neglect subtle signals, including patient appearance or demeanor~\citep{Levin2016, Christ2010, Fernandes2005}.

To address these gaps, we developed a Bayesian framework that integrates prior knowledge and sequentially updates with evolving clinical data~\citep{Gelman2013, Ghosh2019}. As proof of concept, we applied time to first urination (TTU)---an illustrative dynamic, non-verbal cue---within this framework, demonstrated its utility, and developed an application to demonstrate potential deployment in the emergency department (ED). This approach complements rather than replaces existing triage systems and, by extending beyond triage to diagnostic processes, underscores the broader potential of Bayesian models to augment frontline decision-making in acute care.

\section{Results}
\subsection{Study Population}
\begin{figure}[htbp]
  \centering
  \includegraphics[width=0.85\textwidth]{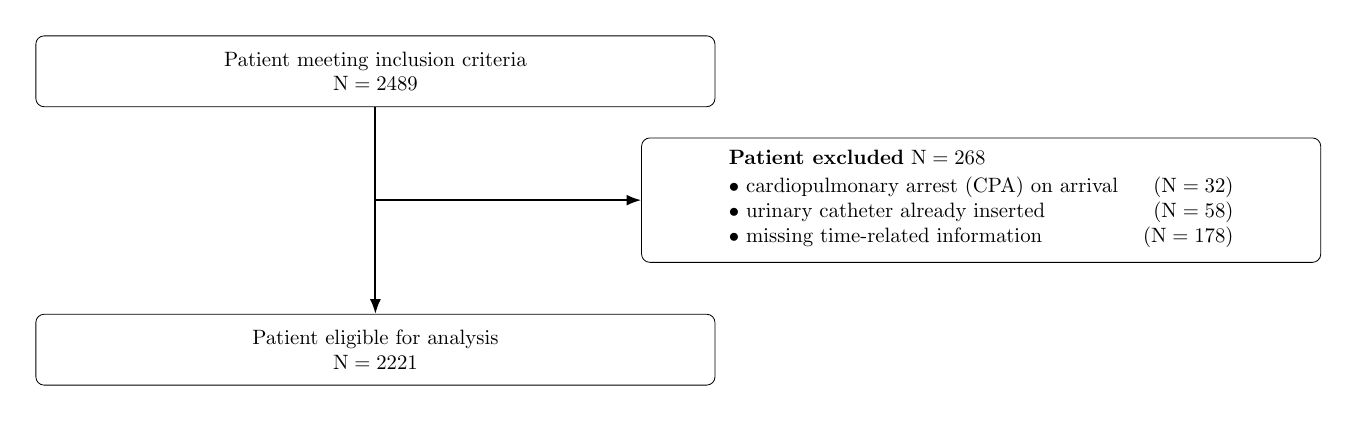}
  \caption{Patient flow.}
  \label{fig:patient-flow}
\end{figure}

Overall, \(n=2{,}489\) ambulance-transported patients met the inclusion criteria during February to August 2025. Following exclusions, the analytic cohort comprised \(n=2{,}221\) patients: \(n=1{,}784\) and \(n=437\) in the model-development and temporal evaluation sets, respectively. Patient flow is presented in \autoref{fig:patient-flow}, and baseline characteristics are summarized in \autoref{tab:demo_dx_combined}.

\begin{table}[htbp]
  \centering
  \caption{Demographics/comorbidities and primary diagnosis categories at ED disposition.}
  \label{tab:demo_dx_combined}
  \small
  \setlength{\tabcolsep}{6pt}
  \begin{minipage}[t]{0.52\linewidth}
    \centering
    \textbf{(A) Demographics and comorbidities}\\[2pt]
    \begin{tabular}{lrr}
      \toprule
      Item & \multicolumn{2}{c}{Value} \\
      \midrule
      Age, median (IQR), years & \multicolumn{2}{c}{74 (51.8--83.0)} \\
      Female sex, $n$ (\%) & \multicolumn{2}{c}{1206 (54.3)} \\
      \midrule
      \multicolumn{3}{l}{\textit{Comorbidities}} \\
      None & \multicolumn{2}{c}{151 (6.8)} \\
      Hypertension & \multicolumn{2}{c}{120 (5.4)} \\
      Diabetes & \multicolumn{2}{c}{61 (2.7)} \\
      Malignancy & \multicolumn{2}{c}{48 (2.2)} \\
      Cerebrovascular disease & \multicolumn{2}{c}{44 (2.0)} \\
      Neurologic disease & \multicolumn{2}{c}{44 (2.0)} \\
      Arrhythmia & \multicolumn{2}{c}{44 (2.0)} \\
      Dyslipidemia & \multicolumn{2}{c}{38 (1.7)} \\
      Heart failure & \multicolumn{2}{c}{31 (1.4)} \\
      Chronic lung disease & \multicolumn{2}{c}{24 (1.1)} \\
      Renal failure & \multicolumn{2}{c}{24 (1.1)} \\
      Psychiatric disorder & \multicolumn{2}{c}{19 (0.9)} \\
      Liver disease & \multicolumn{2}{c}{14 (0.6)} \\
      Valvular disease & \multicolumn{2}{c}{14 (0.6)} \\
      Hematologic disease & \multicolumn{2}{c}{13 (0.6)} \\
      Endocrine disorder & \multicolumn{2}{c}{11 (0.5)} \\
      Peripheral vascular disease & \multicolumn{2}{c}{8 (0.4)} \\
      Connective tissue disease & \multicolumn{2}{c}{7 (0.3)} \\
      Peptic ulcer disease & \multicolumn{2}{c}{5 (0.2)} \\
      \bottomrule
    \end{tabular}
  \end{minipage}\hfill
  \begin{minipage}[t]{0.44\linewidth}
    \centering
    \textbf{(B) Primary diagnosis categories}\\[2pt]
    \begin{tabular}{lr}
      \toprule
      Diagnosis category & $n$ (\%) \\
      \midrule
      Orthopedics (incl.\ fractures) & 117 (5.3) \\
      Gastroenterology & 83 (3.7) \\
      Neurology & 60 (2.7) \\
      Infectious diseases & 47 (2.1) \\
      Cardiovascular & 41 (1.8) \\
      Urology & 28 (1.3) \\
      Cerebrovascular disease & 27 (1.2) \\
      Respiratory & 21 (0.9) \\
      Nephrology & 18 (0.8) \\
      Other & 16 (0.7) \\
      Dermatology & 16 (0.7) \\
      Otolaryngology (ENT) & 11 (0.5) \\
      Psychiatry & 11 (0.5) \\
      Oncology (malignancy) & 9 (0.4) \\
      Endocrinology & 8 (0.4) \\
      None & 6 (0.3) \\
      Hematology & 5 (0.2) \\
      Ophthalmology & 1 (0.0) \\
      Gynecology & 1 (0.0) \\
      \bottomrule
    \end{tabular}
  \end{minipage}
\end{table}

\subsection{Model Convergence and Diagnostics}
Sampling diagnostics indicated stable mixing without pathological divergences, with $\hat{R}\approx 1.00$ for all representative parameters ($\rho_0,\rho_1,\mu_0,\mu_1,\sigma_0,\sigma_1$, and $\beta$ terms). Additionally, trace, rank, and energy plots were visually acceptable in the combined diagnostic bundle (Supplement~1).

\subsection{Admission Rate by TTU}
\begin{figure}[htbp]
  \centering
  \begin{subfigure}[t]{0.48\textwidth}
    \centering
    \includegraphics[width=\linewidth]{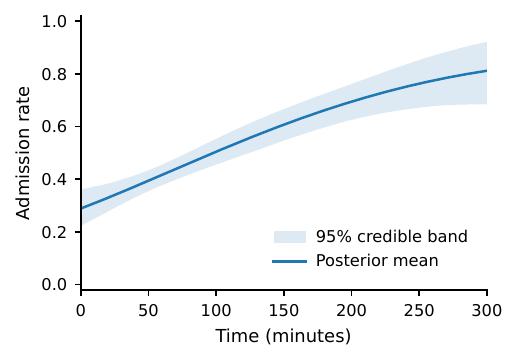}
    \caption{Unadjusted}
    \label{fig:pred-unadj}
  \end{subfigure}\hfill
  \begin{subfigure}[t]{0.48\textwidth}
    \centering
    \includegraphics[width=\linewidth]{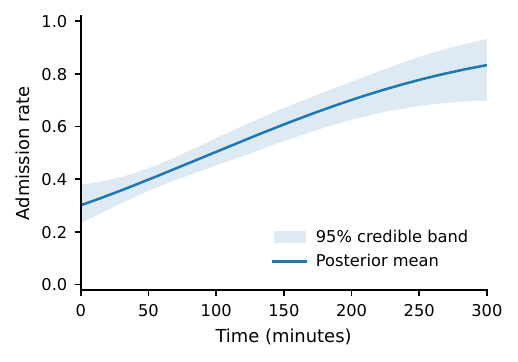}
    \caption{Adjusted for age and sex}
    \label{fig:pred-adj}
  \end{subfigure}
  \caption{Association between post-arrival urination time and hospital admission risk. (a) Unadjusted. (b) Adjusted for age and sex. Posterior mean with 95\% credible interval; estimated via Markov chain Monte Carlo.}
  \label{fig:pred}
\end{figure}

Figure~\ref{fig:pred} displays the predicted relationship between TTU and admission risk for the unadjusted model (panel~a) and the age- and sex-adjusted model (panel~b). The unadjusted model exhibited a marginal association with TTU, whereas the adjusted model provided risk estimates conditional on age and sex. In both panels, (i) admission risk rose with longer TTU, and (ii) this monotonic relationship remained robust after covariate adjustment. Adjustment chiefly shifted the absolute risk level and narrowed the credible bands (i.e., improved precision) while preserving the overall direction. Clinically, TTU is a dynamic bedside cue not explained by basic demographics. A companion application allows users to toggle covariates and observe changes in predicted probabilities and credible bands.\footnote{\url{http://accm.jp/urination/index.html}}

\subsection{Cumulative Admission Proportion and Time-Varying Performance}
\begin{figure}[htbp]
  \centering
  \includegraphics[width=\textwidth]{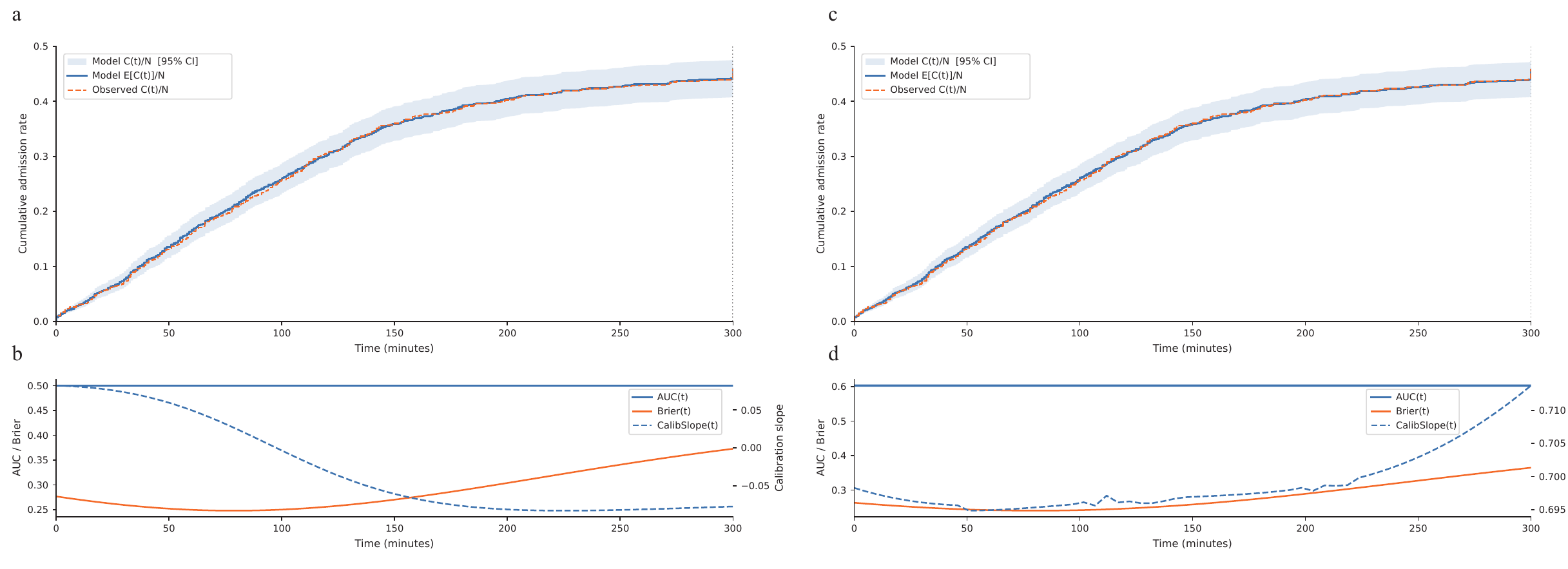}
  \caption{Association between time-to-first urination after ED arrival and hospital admission: covariate-unadjusted cumulative association (a), time-varying performance of the covariate-unadjusted model(b), covariate-adjusted cumulative association (c), and time-varying performance of the covariate-adjusted model(d).}
  \label{fig:merged_figure}
\end{figure}
Cumulative admission proportion in relation to urination time and time-varying performance for the covariate-unadjusted and covariate-adjusted models are provided in \autoref{fig:merged_figure}. Visual inspection demonstrated that both models fit the empirical curves well, and summary metrics in \autoref{tab:gof} corroborated this: area between curves (ABC)/integrated absolute error (IAE), integrated squared error (ISE), root mean square error (RMSE), Kolmogorov–Smirnov (KS), and Cramér–von Mises (CvM) were uniformly low; coverage $\approx 0.98$ and an average 95\% band width $\approx 0.053$ indicated accurate recovery of the cumulative admission trajectory, with credible intervals that contracted where information was abundant and widened at the extremes.

Nevertheless, the two specifications differed in patient-level predictive ability. In the unadjusted model (\autoref{fig:merged_figure} a,b), the time-varying discrimination remained essentially uninformative, with $\mathrm{AUC}(t)\approx 0.5$ throughout, and the calibration slope drifted negatively over time. Accordingly, $\mathrm{Brier}(t)$ was higher ($\approx 0.45$). Contrastingly, in the adjusted model (\autoref{fig:merged_figure} c,d), $\mathrm{AUC}(t)$ improved ($\approx 0.70$), $\mathrm{Brier}(t)$ was low and comparatively stable ($\approx 0.35$), and the calibration slope remained positive ($\approx 0.70$).

Recalibration improved probability alignment without materially changing discrimination; moreover, decision-curve analysis demonstrated modest, threshold-dependent gains for the covariate model, and conclusions were robust to plausible TTU timing error. See Supplement 2 for procedures and full quantitative summaries.

\begin{table}[htbp]
  \centering
  \caption{Goodness of fit for the cumulative curve. Lower values are better for ISE, \(\mathrm{RMSE}_{\text{time}}\), KS, and CvM; empirical coverage near 0.95 is nominal; a narrow average band width is preferred.}
  \label{tab:gof}
  \small
  \setlength{\tabcolsep}{4pt}
  \begin{tabularx}{\linewidth}{*{8}{S[table-format=1.3]} S[table-format=1.2]}
    \toprule
    {Model} &{ISE} & {\(\mathrm{RMSE}_{\text{time}}\)} & {KS} & {CvM} & {ABC} & {IAE} & {Avg.\ band width} & {Coverage (95\%)} \\
    \midrule
    {Unadjusted} & 0.002295 & 0.002766 & 0.008096 & 0.002295 & 0.636472 & 0.636472 & 0.0529 & 0.98 \\
    {Adjusted} & 0.00215 & 0.002766 & 0.008036 & 0.002288 & 0.641434 & 0.637212 & 0.0512 & 0.98 \\
    \bottomrule
  \end{tabularx}
\end{table}
\section{Discussion}
\paragraph{Discussion.}
We introduced a practical approach that updates predictions sequentially as information accrues. Within a Bayesian paradigm, the approach can accommodate established triage systems and severity assessment criteria, combine multimodal diagnostic inputs of differing granularity, and adapt to local practice patterns. Bayesian modeling also provides principled uncertainty quantification and posterior‐predictive calibration diagnostics, which may help address AI hallucination and miscalibration.

We implemented the approach on real‐world clinical data and assessed performance and potential utility. Notably, incorporating the previously overlooked bedside cue TTU contributed an additional signal beyond basic demographics. A demonstration application illustrates potential bedside use.

In the dynamic clinical prediction literature, landmarking and joint modeling (JM) constitute two classical paradigms, while complementary families—including multistate models, state‐space/hidden Markov approaches, time‐varying hazard (varying‐coefficient) models, machine learning–based dynamic survival, and online recalibration frameworks—are increasingly used depending on data‐generating mechanisms and deployment constraints. Landmarking provides time‐consistent risk updates by conditioning on information available at prespecified landmark times; moreover, it has been extended to competing risks and implemented via penalized landmark supermodels and open‐source toolkits~\citep{VanHouwelingen2007, Nicolaie2013, Fries2024penLM, Fries2023dynamicLM}. JM links longitudinal trajectories to the event process to account for measurement error and irregular sampling and is supported by an extensive monograph and Bayesian software (JMbayes)~\citep{Rizopoulos2012Book, Rizopoulos2016JMBayes}. Head‐to‐head comparisons generally frame landmarking and JM as the two mainstream approaches, while hybrid strategies (e.g., landmarking with a Super Learner ensemble) aim to combine their strengths for deployability~\citep{Li2023StatMed, Tanner2021JRSSA}. Notably, each family has implementation‐level limitations. Conversely, our sequential Bayesian updating offers time‐consistent probability updates aligned with information arrival, coherent uncertainty via posterior intervals, hierarchical adaptation to site/time heterogeneity, and natural compatibility with online recalibration, while keeping operational burden modest for ED deployment.

We deliberately selected hospital admission as a “soft” outcome because it permitted high‐density data capture with minimal attrition over short time frames, aligning with prior triage research that has widely modeled admission risk~\citep{Sun2011AEM, Shin2025HIR}. Among existing tools, the emergency severity index has been demonstrated to be both reliable and valid; however, recent multicenter studies report nontrivial mistriage rates and site‐level operational variability, indicating room for real‐world implementation improvement~\citep{Sax2023JNO, Sax2024JamaPed}. These considerations support using admission as the outcome when dynamic bedside cues are layered onto static triage strata. Although incorporating arterial blood gas results and early vital‐sign changes obtained immediately after presentation may yield increased discriminative performance, we intentionally focused on TTU as a broadly applicable, process/behavior–derived signal that is not tied to specific diseases, thereby potentially reflecting global severity. Importantly, previous studies have demonstrated that behavioral observations at first contact (e.g., inability to walk, arrival on a stretcher) strongly predict the need for timely intervention and admission~\citep{Tschoellitsch2023JCM, Riedel2023JCM}. Additionally, given concerns about overreliance on AI decisions and model miscalibration in clinical deployment, our Bayesian framework emphasizes coherent uncertainty and supports online recalibration to maintain well-scaled probabilities under distribution shift~\citep{Feng2022JAMIA}. Finally, because in EHR settings “test not ordered” and measurement delays can themselves be informative~\citep{Tan2023JAMIA}, modeling an easily obtained behavioral cue such as TTU provides a practical countermeasure to informative missingness in conventional diagnostic data streams.

Here, we primarily presented and tested a method that embedded a dynamic diagnostic process via sequential Bayesian updating, rather than a model intended for immediate clinical deployment. At the population level, the model’s fit to the empirical curves was strong; nonetheless, patient‐level performance was modest in the unadjusted specification ($\mathrm{AUC}(t)\approx 0.50$). Following covariate adjustment, discrimination improved ($\mathrm{AUC}(t)\approx 0.70$) and the pattern of negative calibration slopes and elevated $\mathrm{Brier}(t)$ observed in the unadjusted model was attenuated, indicating better‐scaled probabilities. These results are compatible with an independent TTU signal after accounting for age and sex, although interaction effects were not formally tested. To minimize complexity and maintain conceptual focus, we adjusted only for age and sex in addition to TTU. Although incorporating vital signs, chief complaint, and other routinely available features would almost certainly increase discriminative performance, these variables were deliberately omitted at this stage.

This single-center study prioritized feasibility and transparent estimation; nonetheless, prospective external validation is essential to establish generalizability and will be the focus of subsequent work. Although the framework is engineered to interface with existing triage systems and diagnostic pathways, we did not evaluate whether such integration \emph{improves} discrimination or calibration beyond those tools; a head-to-head comparative study is planned as the next step. Despite aligning all predictors to encounter time, residual inclusion of post-decision signals cannot be fully excluded--—a recognized challenge in dynamic prediction that motivates operational safeguards and prospective data-capture audits~\citep{VanHouwelingen2007}. Our primary outcome, hospital admission, reflects both clinical need and system constraints; future studies will include clinically tighter endpoints and composites (e.g., ICU admission, 72-h adverse events, early return). Finally, while the Bayesian design supports online recalibration under distribution shift, we did not implement a formal monitoring/recalibration protocol here; subsequent deployments will specify trigger rules (e.g., calibration drift thresholds) and governance consistent with best-practice recommendations~\citep{VanCalster2019BMCMed}.

Future work should augment the covariate set—including vital‐sign trajectories, chief‐complaint categories, and comorbidity burden—and examine performance when multiple time series and prespecified interactions (e.g., TTU$\times$age/sex/chief complaint) are introduced. Multicenter external validation is warranted to assess transportability, with systematic discrimination and calibration reporting (calibration‐in‐the‐large and slope) across sites and subgroups. Prospective integration studies should test whether adding TTU‐based dynamic updating yields clinically meaningful gains over established triage scores and delineate use‐case boundaries where such gains are realized. Robustness analyses should probe susceptibility to measurement noise, missingness patterns, and process confounding (e.g., catheterization, fluids, and diuretics), alongside time‐consistency audits to mitigate leakage. Finally, evaluation against hard outcomes—including ICU admission, short‐term adverse events, and ED returns—together with decision‐analytic assessments (e.g., net benefit) will be important for threshold selection and clinical adoption.

We introduced and empirically illustrated a practical approach to dynamic clinical prediction in emergency care, using TTU as a widely available behavior/process–derived cue to coherently update probabilities as information accrues. Although not yet deployable, the framework provides a workable foundation for richer, better‐calibrated models that can be recalibrated online and validated across centers prior to routine clinical use.

\section{Methods}
\subsection{Study Design and Setting}
This was a prospective single-center cohort study conducted at Toda Chuo General Hospital, Saitama, Japan. We included all patients transported by ambulance to the ED between February 2025 and August 2025.

\subsection{Data Collection}
During the study period, a structured case report form was completed for each eligible patient, including patient ID, arrival time, clinical outcome, time of first urination complaint, presence/absence of urethral catheter insertion, and name of the recorder. Additional information---chief complaint, vital signs at presentation, and final diagnosis---was abstracted from electronic medical records (EMRs). Disposition (discharge alive, in-hospital death, other) was captured to ensure complete in-hospital follow-up.

\subsection{Participants}
The study included all ambulance-transported patients presenting to the ED between February and August 2025. No a priori power calculation was performed because this proof-of-concept focused on estimating the feasibility of a sequential Bayesian framework. Exclusion criteria were as follows: (1) cardiopulmonary arrest on arrival; (2) a urinary catheter in place at presentation; (3) missing time-related information on the study form. Data from February to June 2025 and July to August 2025 were used for model development and evaluation, respectively.

\subsection{Variables}
Covariates included age, sex, and time from ED arrival to first urination; the primary outcome was hospital admission at ED disposition.

\subsection{Bias}
Selection and information bias were minimized by including all consecutive ambulance arrivals and by using a standardized form and EMR review, respectively. Misclassification of TTU was possible (multi-observer documentation), and residual confounding from unmeasured factors may have remained.

\subsection{Statistical Methods}

\subsubsection{Study Design and Preprocessing}
We analyzed observations \(i=1,\ldots,n\) with a binary outcome \(y_i\in\{0,1\}\) indicating hospital admission at ED disposition. Let \(t_i^{\mathrm{raw}}\ge 0\) (min) denote TTU after ED arrival. Right-censoring was applied at \(C=300\) min, and the censored time was defined as \(t_i=\min\!\bigl(t_i^{\mathrm{raw}},\,C\bigr)\). Moreover, we recorded \(m_i\in\{0,1\}\), an indicator that the patient voided in the ED (uncensored when \(m_i=1\)).

\begin{equation}
t_i=\min\!\bigl(t_i^{\mathrm{raw}},\, C\bigr),\qquad
c_i=\mathbb{I}\!\bigl(t_i^{\mathrm{raw}}>C\bigr).
\end{equation}

Age was standardized, sex was harmonized to a 0/1 indicator, and missingness flags were added. The covariate linear term is given by
\begin{equation}
x_i^\top \beta
= \beta_{\mathrm{age}}\,a_i
+ \beta_{\mathrm{age\text{-}mis}}\,a^{\mathrm{mis}}_i
+ \beta_{\mathrm{sex}}\,\mathrm{sex01}_i
+ \beta_{\mathrm{sex\text{-}mis}}\,s^{\mathrm{mis}}_i.
\end{equation}

\subsubsection{Bayesian Hierarchical Model}
We introduced baseline event propensities $\rho_0,\rho_1\in(0,1)$, constrained to $(\varepsilon,1-\varepsilon)$ with $\varepsilon=10^{-6}$. Two normal kernels with ordered means $\mu_1\ge\mu_0$ :
\begin{equation}
f_0(t)=\phi(t\mid\mu_0,\sigma_0), \qquad
f_1(t)=\phi(t\mid\mu_1,\sigma_1),
\end{equation}
where $\phi$ is the normal density, and $\overline{\Phi}$ its survival function. Standard deviations were parameterized as follows:
\begin{equation}
\sigma_k=\exp(\log\sigma_k)+\varepsilon,\qquad k\in\{0,1\}.
\end{equation}

\subsubsection{Likelihood Specification}
For informative times ($m_i=1$), the effective log-likelihood ratio with censoring is given by
\begin{equation}
\Delta_i =
(1-c_i)\{\log f_1(t_i)-\log f_0(t_i)\}
+ c_i\{\log \overline{\Phi}((C-\mu_1)/\sigma_1)
- \log \overline{\Phi}((C-\mu_0)/\sigma_0)\}.
\end{equation}

The log-odds for observation $i$ is
\begin{equation}
\mathrm{logit}(p_i)
= (1-m_i)\!\left[\log\frac{1-\rho_1}{1-\rho_0}+x_i^\top\beta\right]
+ m_i\!\left[\log\frac{\rho_1}{\rho_0}+\Delta_i+x_i^\top\beta\right].
\end{equation}

Accumulating posterior $\mathrm{logit}_i(t)$ samples yields the posterior cumulative incidence:
\begin{equation}
\hat{C}(t)=\frac{1}{N_1}\sum_{i=1}^{N_1}\mathbb{E}\bigl[p_i(t)\bigr],
\end{equation}
where $N_1$ denotes the number of individuals with $m=1$. The outcome model is given by
\begin{equation}
y_i\sim\mathrm{Bernoulli}(p_i), \qquad
p_i=\mathrm{logit}^{-1}\!\bigl(\mathrm{logit}(p_i)\bigr).
\end{equation}

\subsubsection{Priors}
Weakly informative priors were employed as follows:
\begin{itemize}
  \item $\rho_{k}=\operatorname{logit}^{-1}(\eta_{k}),\ \eta_{k}\sim\mathcal{N}(0,1),\quad k\in\{0,1\}$ 
  \item $\mu_0\sim \mathrm{TruncatedNormal}\!\bigl(t_{\mathrm{mean}},\,2\cdot \mathrm{scale},\,0,\,C\bigr)$
  \item $\mu_1=\min\!\bigl(\mu_0+\delta_\mu,\,C\bigr),\quad \delta_\mu\sim \mathrm{HalfNormal}(\mathrm{scale})$
  \item $\log\sigma_k\sim\mathcal{N}\!\bigl(\log(\mathrm{scale}),\,1\bigr),\quad k\in\{0,1\}$
  \item $\beta_\ast \sim \mathcal{N}(0,1)\ \text{independently}$
\end{itemize}

\noindent where $\mathrm{scale}\coloneqq \max\{\operatorname{sd}(t),\,1.0\}$ (min) with $t_j=\min(t^{\mathrm{raw}}_j,\,C)$.

\subsubsection{Inference and Diagnostics}
For sampling, we used the No-U-Turn Sampler (NUTS) with four chains, a warm-up of 3{,}000 iterations per chain, and 3{,}000 posterior draws per chain. Moreover, we set \texttt{target\_accept} = 0.90 and \texttt{max\_treedepth} = 12. Convergence was assessed via $\hat{R}$, effective sample size, Monte Carlo standard error, and visual diagnostics (trace, rank, energy).

\subsubsection{Model Evaluation (Population- and Patient-Level)}
\paragraph{Population-level fit of cumulative admissions.}

We compared the observed cumulative proportion \(C_{\mathrm{obs}}(t)/N\) with the model-based posterior mean \(E[C(t)]/N\) over \(t\in[0,300]\) min, reporting pointwise 95\% credible bands. Furthermore, we summarized time-integrated discrepancy using the following metrics: ABC, IAE, ISE, time-domain RMSE (\(\mathrm{RMSE}_{\text{time}}\)), KS (sup-norm), CvM, empirical coverage of the pointwise 95\% bands, and average band width.

\paragraph{Individual-level metrics at landmark times.}
At \(\mathcal{T}=\{60,120,180,240,300\}\) min, we computed per-patient predicted probabilities \(\hat p_i(t)\) and outcomes \(y_i(t)\); moreover, we evaluated (i) discrimination in terms of \(\mathrm{AUC}(t)\), i.e., the probability that a case has a higher \(\hat p(t)\) compared with that of a non-case; (ii) accuracy via \(\mathrm{Brier}(t)=\frac{1}{n_t}\sum_{i=1}^{n_t}\bigl(\hat p_i(t)-y_i(t)\bigr)^2\); (iii) calibration at time \(t\) using the intercept from \(y_i(t)\sim \alpha\) (calibration-in-the-large) and the slope \(\beta\) from \(y_i(t)\sim \alpha+\beta\,\mathrm{logit}\!\bigl(\hat p_i(t)\bigr)\) (ideal: intercept \(=0\), slope \(=1\)), and (iv) the 10-bin expected calibration error \(\mathrm{ECE}_{10}(t)\).

\paragraph{Recalibration.}
At each $t$, logistic recalibration was applied as follows:
\begin{equation}
\hat p_i^{\,\star}(t)=\mathrm{logistic}\!\bigl(\hat\alpha(t)+\hat\beta(t)\,\mathrm{logit}(\hat p_i(t))\bigr).
\end{equation}
Additionally, we reported pre-/post-hoc intercepts and slopes on held-out data. Full procedures and figures are provided in Supplement 2.

\subsection{Ethics Approval and Consent to Participate}
The study adhered to the Declaration of Helsinki and was approved by the Institutional Review Board of Toda Chuo General Hospital (Approval No.\ 0644). Given the observational, de-identified design, the need for informed consent was waived.

\section{Data Availability}
The minimal dataset supporting the findings is available upon reasonable request from the corresponding author.
\section{Code Availability}
Sample analysis code and the bedside demonstration app are available at https://github.com/atsushi-tmdu/TTU
and also distributed on the App Store (https://apps.apple.com/us/app/urination-time-triage/id6752229741). We recommend citing a release tag/commit for reproducibility.

\section{Author Contributions}
AS: Conceptualization;  Investigation; Methodology; Software; Formal analysis; Visualization; Writing—original draft; Project administration; Supervision.
YT: Investigation; Methodology; Validation; Data curation \& editing.
RI: Supervision; Writing—review \& editing.
HS: Investigation; Data curation; Writing—review \&
AE: Investigation; Resources; Data curation; Writing—review \& editing.
KM: Supervision;  Funding acquisition \& editing.

\section{Competing interests}
The authors declare no competing interests.
\bibliographystyle{naturemag} 
\bibliography{refs}

\clearpage
\appendix
\section*{Supplement 1: Sampling diagnostics (NUTS/HMC)}
\addcontentsline{toc}{section}{Supplement 1: Sampling diagnostics (NUTS/HMC)}
\subsection*{S1.A — Energy Diagnostics}
Hamiltonian Monte Carlo sampling was evaluated using the energy plot
, which displays (i) the marginal distribution of the Hamiltonian energy $E$ and (ii) the distribution of energy transitions $\Delta E$ between successive proposals. Adequate overlap and dispersion between these two distributions indicate efficient exploration of the typical set. As a scalar summary, we report the energy Bayesian fraction of missing information (E\,-BFMI); values $\ge 0.3$ are generally considered adequate, whereas low E\,-BFMI suggests poor energy exploration and potential pathologies (e.g., sticky trajectories or a need for reparameterization).

\begin{figure}[htbp]
  \centering
  \includegraphics[width=0.85\textwidth]{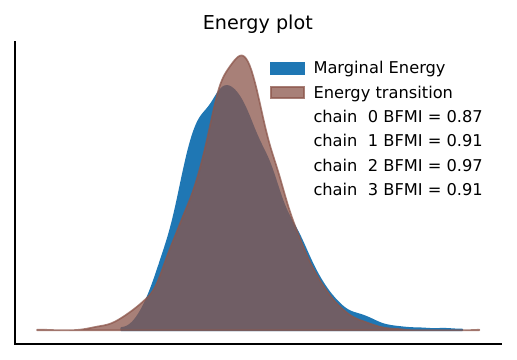}
  \captionsetup{labelformat=empty} 
  \caption*{\textbf{Figure S1A.} Energy plot with E\,-BFMI by chain.}
\end{figure}
\clearpage


\FloatBarrier                 
\needspace{0.8\textheight}    

\subsection*{S1.B — Trace Diagnostics (By Chain)}
We inspected trace plots for representative parameters $(\rho_0,\rho_1,\mu_0,\mu_1,\sigma_0,\sigma_1,\boldsymbol{\beta})$ across chains to assess mixing and stationarity under NUTS/HMC. Well-behaved traces demonstrate rapid mixing, “fat-caterpillar” shapes without persistent trends, and stable running means after warm-up. Between-chain overlays were visually indistinguishable, and autocorrelation decayed rapidly, consistent with convergence and adequate effective sample sizes.
\begin{figure}[!htbp]
  \centering
  \includegraphics[width=\linewidth,height=0.70\textheight,keepaspectratio]{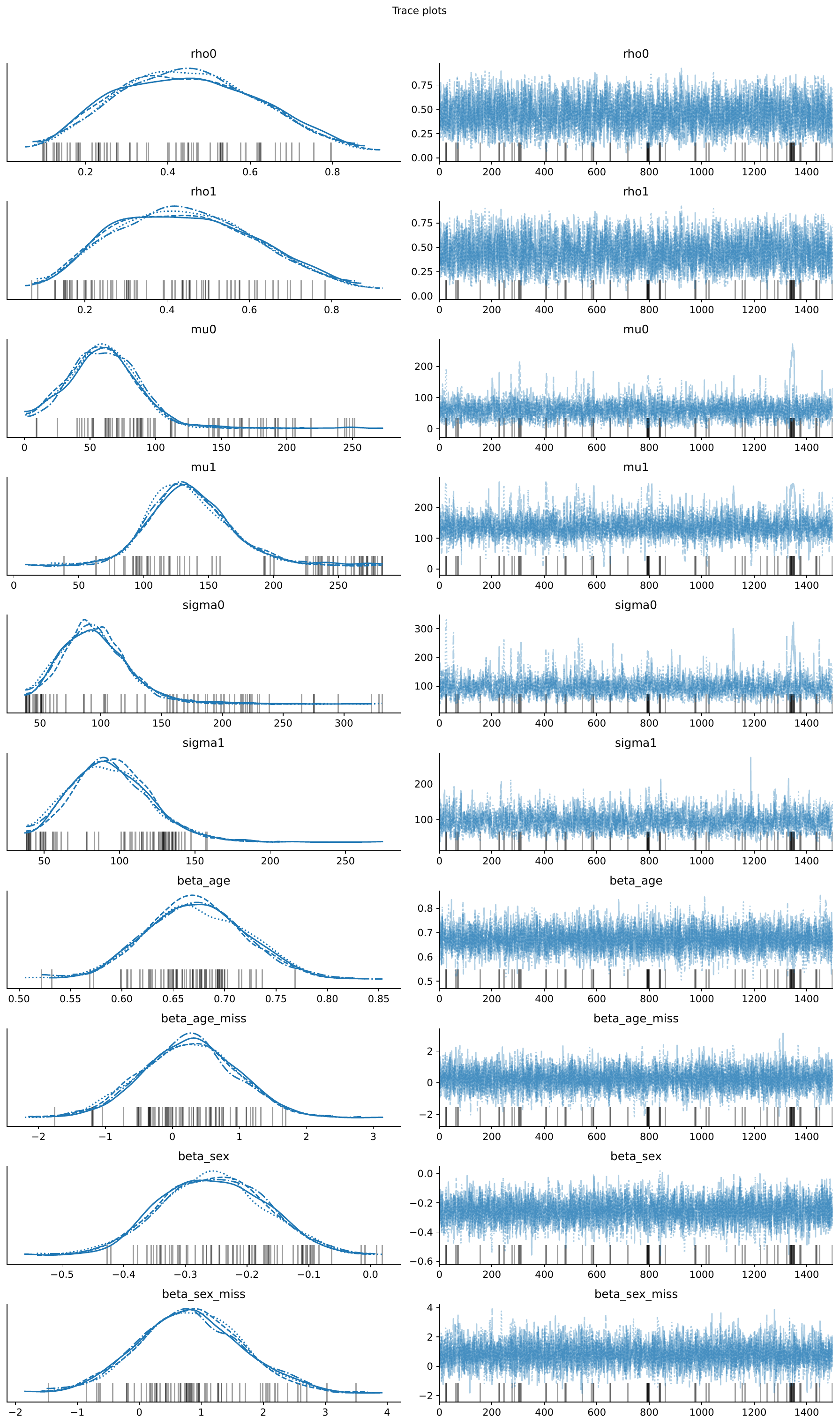}
  \caption*{\textbf{Figure S1B.} Trace plots by chain with running means and autocorrelation panels.}
\end{figure}

\clearpage

\subsection*{S1.C —Rank Diagnostics (Uniform Rank Histograms)}
Convergence was further evaluated with rank plots (uniform rank histograms). For each parameter, draws from each chain are ranked within the pooled posterior; under good mixing, per-chain rank frequencies are approximately uniform. Systematic deviations (e.g., U- or $\cap$-shapes or edge spikes) indicate poor between-chain overlap or adaptation issues. The observed histograms were approximately flat with only random fluctuation, supporting convergence.

\begin{figure}[htbp]
  \centering
  \includegraphics[width=0.95\textwidth]{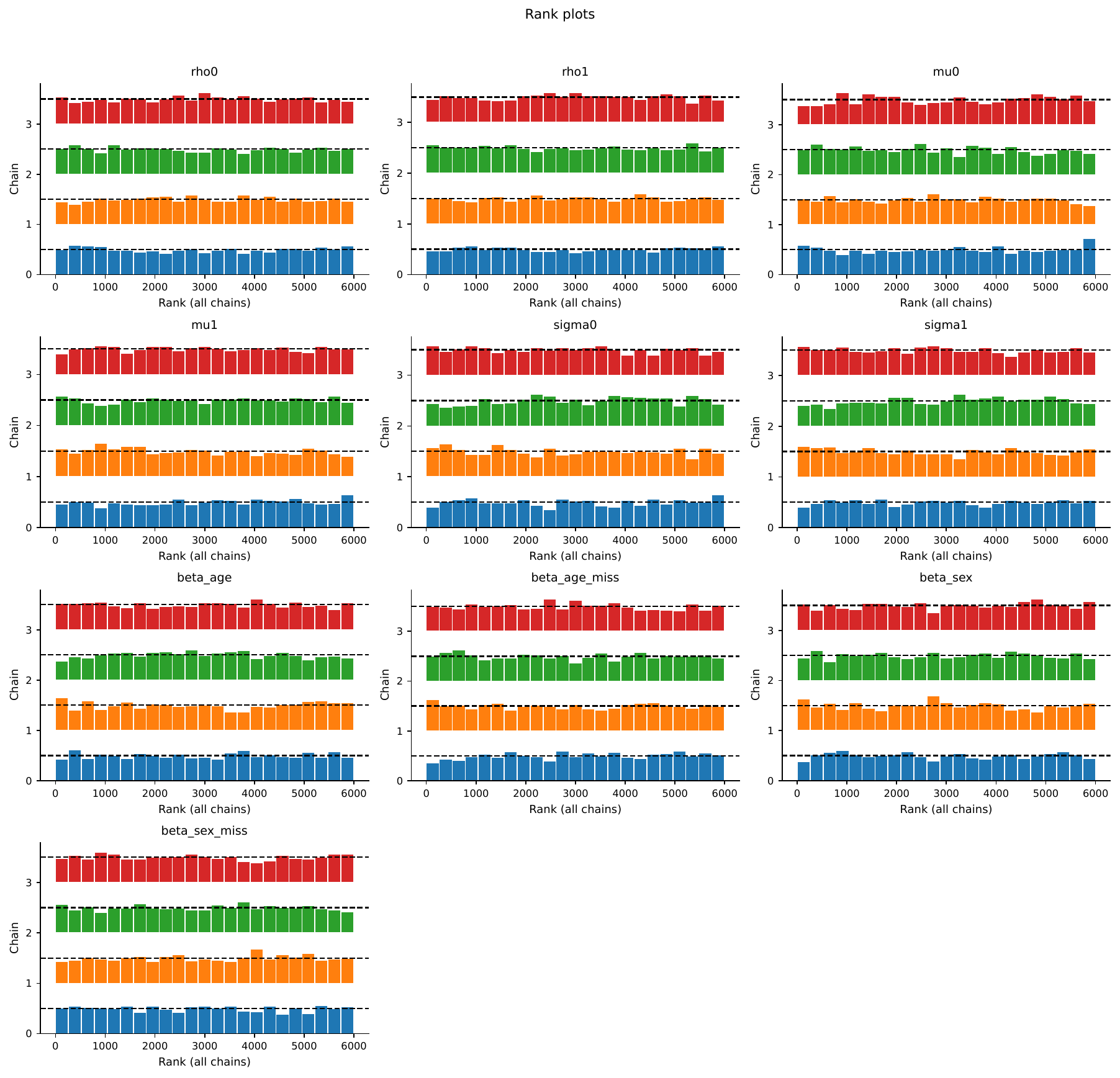}
  \caption*{\textbf{Figure S1C.} Uniform rank histograms by chain.}
\end{figure}

\clearpage

\section*{Supplement 2: Recalibration, Decision Curves, and Robustness}
\addcontentsline{toc}{section}{Supplement 2: Recalibration, Decision Curves, and Robustness}

\subsection*{S2.A — Logistic/Platt Recalibration (One Figure)}
\addcontentsline{toc}{subsection}{S2.A — Logistic/Platt recalibration}

\noindent\textbf{Model.}\;
$\operatorname{logit}(p^\star)=\alpha(t)+\beta(t)\operatorname{logit}(\hat p)$ at each landmark $t$.

\noindent\textbf{Panel A.}\; Calibration at $t=120$ and $t=300$: observed event rate (deciles) vs.\ $\hat p$; overlay pre-/post-Platt lines (ideal: slope $=1$, intercept $=0$).\\
\textbf{Panel B.}\; Distribution shift: histogram/density of $\hat p$ vs.\ $p^\star$; annotate mean absolute change and \% crossing thresholds (e.g., $0.2/0.4$).

\begin{figure}[htbp]
  \centering
  \begin{minipage}{0.48\linewidth}
    \centering
    \includegraphics[width=\linewidth]{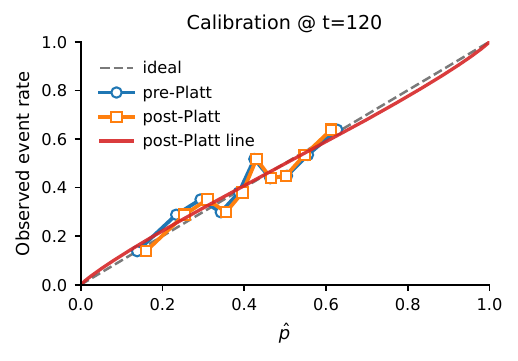}
    \caption*{S2.A — Panel A (t=120)}
  \end{minipage}\hfill
  \begin{minipage}{0.48\linewidth}
    \centering
    \includegraphics[width=\linewidth]{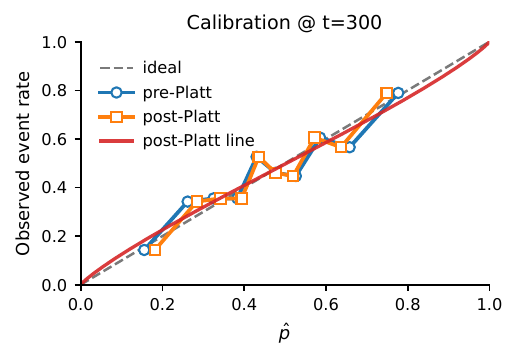}
    \caption*{S2.A — Panel A (t=300)}
  \end{minipage}

  \medskip

  \begin{minipage}{0.48\linewidth}
    \centering
    \includegraphics[width=\linewidth]{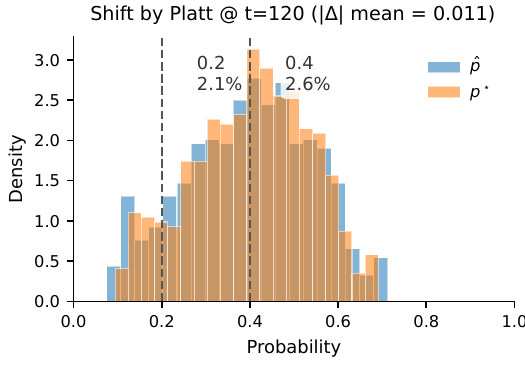} 
    \caption*{S2.A — Panel B (t=120)}
  \end{minipage}\hfill
  \begin{minipage}{0.48\linewidth}
    \centering
    \includegraphics[width=\linewidth]{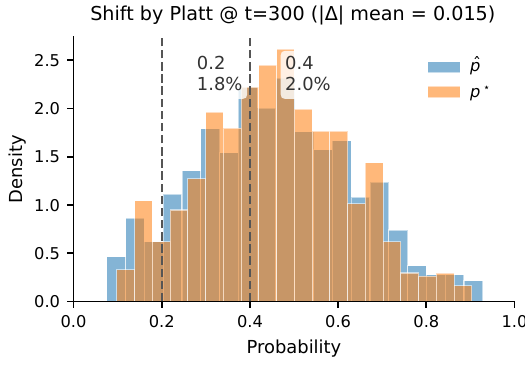}
    \caption*{S2.A — Panel B (t=300)}
  \end{minipage}

  \caption{S2.A — Platt recalibration at selected landmarks ($t=120, 300$).}
  \label{fig:s2-platt}
\end{figure}

\noindent\textbf{Findings from our run.}\;
Platt recalibration at $t=120/300$ improved calibration toward the 45° line
($\alpha_{120}=-0.010,\ \beta_{120}=0.899$; $\alpha_{300}=-0.013,\ \beta_{300}=0.879$; 95\% CrIs: add).
Discrimination was unchanged (AUC$_{120}$ 0.655$\to$0.655; AUC$_{300}$ 0.681$\to$0.681).
The mean absolute shift $|\Delta|$ was 0.011 ($t=120$) and 0.015 ($t=300$);
only 2.6\% and 2.0\% crossed $p_t=0.4$, respectively—indicating improved probability
interpretability with minimal decision impact.

\bigskip

\subsection*{S2.B — Decision-Curve Analysis}
\addcontentsline{toc}{subsection}{S2.B — Decision-Curve Analysis}

\noindent\textbf{Purpose.}\; Quantify clinical net benefit across threshold probabilities $p_t$.\\
\textbf{Panel.}\; Curves for Baseline (time only), Covariate (time+age+sex), \textit{Admit all}, \textit{Admit none} over $p_t\in[0.1,0.6]$.\\
\textbf{Formula.}\;
$\displaystyle \mathrm{NB}(p_t)=\frac{\mathrm{TP}}{N}-\frac{\mathrm{FP}}{N}\frac{p_t}{1-p_t}$.

\begin{figure}[htbp]
  \centering
  \includegraphics[width=0.9\textwidth]{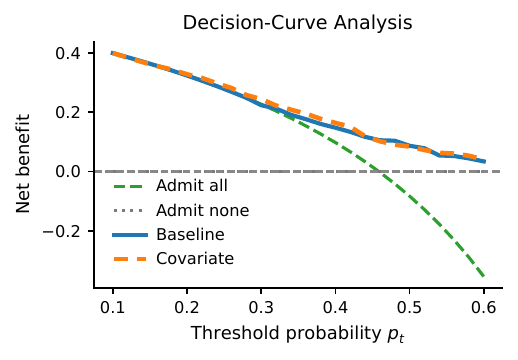}
  \caption{S2.B — Decision-curve analysis comparing Baseline vs.\ Covariate with \textit{Admit all}/\textit{Admit none} references.}
  \label{Fig:s2-dca}
\end{figure}

\noindent\textbf{Reporting.}\; $\Delta\mathrm{NB}$ (Covariate $-$ Baseline):
$p_t=0.2$:0.0043 (95\% CI $-$0.0848,0.0887);
$p_t=0.3$:0.0205 (95\% CI $-$0.0779,0.1151);
$p_t=0.4$:0.0162 (95\% CI $-$0.0767,0.1060).
Interpretation: Covariate shows small, directionally favorable net benefit for $p_t=0.2$–$0.4$, but CIs include zero.

\bigskip
\clearpage
\subsection*{S2.C — Robustness to TTU Measurement Error (One Table)}
\addcontentsline{toc}{subsection}{S2.C — Robustness to TTU measurement error}

\noindent Add jitter to $t_i^{\mathrm{raw}}$: $\tilde t_i = t_i^{\mathrm{raw}} + \epsilon$, with $\epsilon\sim\mathrm{Uniform}(-\delta,\delta)$, $\delta\in\{5,10\}$ min; then recompute landmark metrics from posterior predictive draws.

\begin{table}[htbp]
  \centering
  \caption{S2.C — Robustness of time-dependent metrics to TTU jitter (landmarks in minutes).}
  \label{tab:s2-robust-jitter}
  \small
  \setlength{\tabcolsep}{3pt}
  \begin{tabular}{rccccc ccccc ccc ccc}
    \toprule
    \multicolumn{1}{c}{Jitter (±min)} &
    \multicolumn{5}{c}{AUC} & \multicolumn{5}{c}{Brier} &
    \multicolumn{2}{c}{Cal.\ Intercept} &
    \multicolumn{2}{c}{Cal.\ Slope} \\
    \cmidrule(lr){2-6} \cmidrule(lr){7-11} \cmidrule(lr){12-13} \cmidrule(lr){14-15}
    & 60 & 120 & 180 & 240 & 300 & 60 & 120 & 180 & 240 & 300 & 120 & 300 & 120 & 300 \\
    \midrule
    0.0  & 0.642 & 0.655 & 0.651 & 0.662 & 0.681 & 0.211 & 0.223 & 0.228 & 0.228 & 0.223 & -0.010 & -0.010 & 0.899 & 0.884 \\
    5.0  & 0.642 & 0.650 & 0.648 & 0.660 & 0.680 & 0.212 & 0.224 & 0.229 & 0.228 & 0.223 & -0.036 & -0.011 & 0.861 & 0.876 \\
    10.0 & 0.642 & 0.650 & 0.651 & 0.662 & 0.682 & 0.210 & 0.224 & 0.228 & 0.227 & 0.222 & -0.045 & -0.009 & 0.870 & 0.891 \\
    \bottomrule
  \end{tabular}
\end{table}

\noindent\textbf{One-liner for Results.}\;
Sensitivity with $\pm 5/\pm 10$-minute TTU jitter yielded $\Delta\mathrm{AUC}(t)\le 0.005$, $\Delta\mathrm{Brier}(t)\le 0.001$, and $\Delta$Calibration slope $\le 0.038$ across landmarks; qualitative conclusions were preserved.

\end{document}